\documentclass[journal=jctccc,manuscript=article]{achemso}

\usepackage[version=3]{mhchem} 

\author{J\'{o}gvan Magnus Haugaard Olsen}
\email{jmo@sdu.dk}
\affiliation[University of Southern Denmark]
{Department of Physics, Chemistry and Pharmacy, University of Southern Denmark, Campusvej 55, DK-5230 Odense M, Denmark.}
\author{Erik Donovan Hedeg\aa{}rd}
\email{erik.hedegard@teokem.lu.se}
\affiliation[Lund University]
{Department of Chemistry, Lund University, Kemicentrum S{\"o}lvegatan 39, Lund, Sweden.}

\title{Modeling the absorption spectrum of the permanganate ion in vacuum and in aqueous solution}

\begin{document}

\begin{abstract}
The absorption spectrum of the \ce{MnO4-} ion has been a test-bed for quantum-chemical methods over the last decades.
Its correct description requires highly-correlated multiconfigurational methods, which are incompatible with the inclusion of finite-temperature and solvent effects due to their high  computational demands.
Therefore, implicit solvent models are usually employed.
Here we show that implicit solvent models are not sufficiently accurate to model the solvent shift of \ce{MnO4-}, and we analyze the origins of their failure.
We obtain the correct solvent shift for \ce{MnO4-} in aqueous solution by employing the polarizable embedding (PE) model combined with a range-separated complete active space short-range density functional theory method (CAS-srDFT).
Finite-temperature effects are taken into account by averaging over structures obtained from ab initio molecular dynamics simulations.
The explicit treatment of finite-temperature and solvent effects facilitates the interpretation of the bands in the low-energy region of the \ce{MnO4-} absorption spectrum, whose assignment has been elusive.
\end{abstract}

\section{Introduction}
Inclusion of environment effects in computational chemistry is pivotal to achieve the best possible match between theoretical models and experimental measurements. The presence of an environment influences both energetics and spectroscopic constants. An example of the latter  is how the presence of a molecular environment can alter 
absorption spectra, and many chemical compounds have a characteristic shift of their excitation energies that depends on the specific environment.
For solutes, the \textit{solvent shift} may be defined as $\Delta E_{\mathrm{shift}} = E_{\mathrm{sol}} -  E_{\mathrm{vac}}$, where $E_{\mathrm{sol}}$ and $E_{\mathrm{vac}}$ are excitation energies obtained with and without solvent included, respectively.
Accurate prediction of excitation energies and solvent shifts poses special demands on the theoretical model. 
Obviously it is crucial that the theoretical method is able to describe the electronic structure to high accuracy.
For a solvated compound, it is also important to accurately model the environment-induced perturbation of the solute's charge density. 
Furthermore, it is necessary to include finite-temperature effects which can be done by averaging over several structures from a molecular dynamics (MD) trajectory.
Transition metal complexes can generally be expected to display multireference character, and in such cases both dynamical and static correlation must be modeled accurately with multireference methods, such as complete active space second-order perturbation theory (CASPT2).\cite{Andersson1992}
A prime example is the \ce{MnO4-} ion whose absorption spectrum has been a test-bed for quantum-chemical methods\cite{wolfsberg1952,Johnson_1971,Johansen_1983,Nakai_1991,Nooijen_1999,Gisbergen_1999,Nooijen_2000,Boulet_2001,Neugebauer_2005,Linta_2012,Ziegler_2012,Almeida_2015,Seidu_2015} for decades.
Despite its apparent simplicity, the closed-shell (d$^0$) \ce{MnO4-} ion has been shown to have sizable non-dynamical correlation.\cite{Buijse_1990}
Ab initio methods based on a single-reference wave function have resulted in different assignments\cite{Johansen_1983,Nakai_1991,Nooijen_1999,Nooijen_2000,Almeida_2015} of the experimental spectrum\cite{Holt_1967}. 
The ubiquitous time-dependent density functional theory (TD-DFT) has also been applied, but is noted to significantly overestimate the absolute excitation energies\cite{Gisbergen_1999,Boulet_2001,Neugebauer_2005}, although recent developments by Ziegler and coworkers seems to offer some improvement\cite{Seidu_2015}.
An obstacle for the application of multireference methods has been that the required active space has been estimated to be 24 electrons in 17 orbitals, which is rather large.\cite{Veryazov_2011,Hedegaard_2014,Stein_2016}
A recent study of the electronic absorption spectrum managed to reach the required active space with a restricted active space method (RASPT2).\cite{Su_2013} 
Although good agreement with experiment was obtained for the intense transitions, there are still aspects of the \ce{MnO4-} absorption spectrum that remains puzzling. Transitions in the low-energy region are still not assigned\cite{Holt_1967,Houm_ller_2013}, and with the measurement of accurate gas-phase data\cite{Houm_ller_2013}, experimental estimates of the vacuum to aqueous solvent shift for the first and third electronic bands are available; both are small blue-shifts. Yet previous theoretical studies employing continuum solvation models have predicted either a negligible solvent shift of $0.01-0.02$ eV\cite{Su_2013,Seidu_2015} or a somewhat more pronounced red-shift of $-0.13$ eV\cite{Neugebauer_2005,Seth_2004}, in clear contrast to experiment.
It is presently not known whether the discrepancies are caused by inadequate electronic structure methods, solvent models, or both.

In this paper, we address the theoretical description of the \ce{MnO4-} absorption spectrum in both vacuum and aqueous solvent. To ensure an accurate description it seems that methods that can capture multiconfigurational character are required. Here we use a method that combines Kohn-Sham (KS) density functional theory (DFT) with a multiconfigurational wave function.
A number of such schemes are currently in development \cite{savinbook,Savin_1995,leininger1997,Angyan_2005,Fromager_2007,Fromager_2008,Fromager_2010,Toulouse_2009,Fromager_2013,Hedegaard_2013b,Hedegaard_2015b,Hedegaard_2016a,Hubert_2016a,Hubert_2016b,Senjean_2016,Grimme_1999,Marian_2008,Manni_2014,Garza_2015} and here we employ a range-separation method denoted complete active space short-range DFT (CAS-srDFT).\cite{Fromager_2007,Fromager_2008,Fromager_2013,Hedegaard_2013b,Hedegaard_2015b,Hubert_2016a}
The CAS-srDFT method is computationally cheaper than perturbation-based multireference methods and has been shown to be of comparable accuracy\cite{Hubert_2016a,Hubert_2016b,Hedegaard_2016a}.
Further, its simultaneous treatment of dynamical and static correlation enables the use of significantly smaller active spaces without much loss in accuracy. 
We exploit this here to go beyond what have been done so far, and also consider finite-temperature effects as well as the solvent effects on the electronic absorption spectrum.
Thus, we calculate CAS-srDFT excitation energies and oscillator strengths using a series of structures taken from ab initio MD and 
combined quantum mechanics / molecular mechanics (QM/MM) ab initio MD simulations.
From these underlying structures we also show that continuum solvation models describe the electrostatic perturbation by the solvent inadequately.
We will instead employ the polarizable embedding (PE) model\cite{Olsen_2010,Olsen_2011} which uses an advanced, classical potential to model the electrostatic solvent effects.
With the combined use of QM/MM MD and PE-CAS-srDFT\cite{Hedegaard_2015a}, we are able to reproduce the experimentally-obtained blue shift of the lowest intense transition.
Furthermore, we assign the low-energy parts of the \ce{MnO4-} absorption spectrum.

\section{Results and discussion}

First we investigate the performance of the CAS-srDFT method on the \ce{MnO4-} ion in vacuum and in aqueous solvent described by the polarizable continuum model (PCM).
The four lowest intense transitions in T$_d$ symmetry (\textit{i.e.}~transitions to the T$_2$ states) are examined.
We use symmetric, geometry-optimized structures, \textit{i.e.}, we neglect finite-temperature effects.
We performed the calculations based on two geometries: a vacuum geometry (BLYP/6-31G) and a solvent (PCM) geometry (PCM-BLYP/6-31G).
The lowest vertical excitation energies, as well as the associated oscillator strengths, were obtained for both structures based on time-dependent CAS(14,12)-srPBE calculations with and without PCM embedding.
This allows us to separate the solvent effects on the spectrum into a contribution stemming from the direct electrostatic interactions that polarizes the electron density of the solute, and an indirect contribution which is caused by the solvent-induced change in molecular geometry.

\begin{table}
	\small
	\caption{Vertical excitation energies in eV and oscillator strengths of the lowest intense transitions of the \ce{MnO4-} ion in vacuum or aqueous solution calculated using geometry-optimized structures}
	\label{tbl:geomopt}
	\resizebox{\columnwidth}{!}{
	\begin{tabular}{cccccccccccccc}
		\hline
		Method & \multicolumn{4}{c}{CAS(14,12)-srPBE$^{a}$} & \multicolumn{4}{c}{PCM$^{b}$-CAS(14,12)-srPBE$^{a}$} & \multicolumn{2}{c}{RAS(24,17)PT2$\cite{Su_2013}$} & \multicolumn{3}{c}{experiment$^c$} \\
		Geom. & \multicolumn{2}{c}{vacuum} & \multicolumn{2}{c}{PCM$^b$} & \multicolumn{2}{c}{vacuum} & \multicolumn{2}{c}{PCM$^b$} & \multicolumn{2}{c}{vacuum} & vacuum\cite{Houm_ller_2013} & aq. sol.\cite{Houm_ller_2013} & crys.\cite{Holt_1967} \\
		\hline
		State & $E_{\mathrm{vac}}$ & $f$ & $E_{\mathrm{vac}}$ & $f$ & $E_{\mathrm{aq}}$ & $f$ & $E_{\mathrm{aq}}$ & $f$ & $ E_{\mathrm{vac}}$ & $f$ & $E_{\mathrm{vac}}$ & $E_{\mathrm{aq}}$ & $E_{\mathrm{xtal}}$ \\
		\hline
		$1^{1}\mathrm{T}_{2}$ & 2.21 & 0.095 & 2.25 & 0.092 & 2.16 & 0.127 & 2.20 & 0.123 & 2.33 & 0.004 & 2.09/2.18 & 2.27/2.35 & 2.3/2.4 \\
		$2^{1}\mathrm{T}_{2}$ & 3.61 & 0.010 & 3.64 & 0.010 & 3.61 & 0.018 & 3.64 & 0.018 & 3.53 & 0.002 & 3.55 & - & 3.6 \\
		$3^{1}\mathrm{T}_{2}$ & 4.01 & 0.084 & 4.05 & 0.082 &  -   &   -   &  -   &   -   & 4.20 & 0.006 & 3.94 & 4.00 & 4.1 \\
		$4^{1}\mathrm{T}_{2}$ & 5.13 & 0.171 & 5.19 & 0.162 &  -   &   -   &  -   &   -   & 5.72 & 0.002 & - & - & 5.4 \\
		\hline
	\end{tabular}} \\
	$^a$ \footnotesize{Calculated using a development version of Dalton\cite{daltonpaper,dalton_version} with cc-pVDZ basis set and the srPBE(HSE;RI) exchange-correlation functional\cite{Heyd_2003}, as defined in ref.~\citenum{Fromager_2007}.}
	$^b$ \footnotesize{Cavity was built using 1.9 and 1.52 {\AA} for Mn and O atomic radii, respectively.
		$^c$ Estimated absorption maxima from refs.~\citenum{Houm_ller_2013} and \citenum{Holt_1967}.}
\end{table}

The results for the four lowest intense transitions are shown in table~\ref{tbl:geomopt}.
The CAS(14,12)-srPBE calculation based on the vacuum geometry gives the values that can be compared to the accurate results based on RAS(24,17)PT2.\cite{Su_2013}
In terms of excitation energies both methods agree to within 0.1 eV for the first two transitions, and 0.2 eV for the third transition.
However, there is a relatively large discrepancy for the fourth transition where they differ by 0.6 eV.
We note that \citeauthor{Su_2013}\cite{Su_2013} actually report two states with low oscillator strength below $4^{1}\mathrm{T}_{2}$, and the state that we have designated as $4^{1}\mathrm{T}_{2}$ is actually $6^{1}\mathrm{T}_{2}$ in ref.~\citenum{Su_2013}.
The oscillator strengths are rather different though the same trends are observed except for the fourth state.
This is not unprecedented, as we recently showed that CASPT2 and CAS-srDFT can obtain oscillator strengths that differ significantly\cite{Hedegaard_2016a}, although they usually predict the same trends.
Some of the differences have been attributed to the fact that oscillator strengths in CASPT2 (and RASPT2) methods are effectively obtained at the CASSCF level\cite{Hedegaard_2016a}.
For these higher-lying excitations (close to the ionization limit) the inclusion of Rydberg basis sets can be of importance, although it had only little effect in ref.~\citenum{Su_2013}.
CAS-srDFT has previously been noted to be more sensitive than \textit{e.g.}~CASPT2 to Rydberg-valence mixing\cite{Hubert_2016a,Hubert_2016b} which could also be a possible explanation for the differences.
CAS(14,12)-srPBE compares very well with experimental values for the first three excitations for which there exist experimental values.
Although there are no experimental values for the fourth transition in vacuum it has been measured in crystalline phase, and comparing to this value it seems that CAS(14,12)-srPBE slightly underestimates the excitation energy whereas RAS(24,17)PT2 overestimates it by about the same amount.

The most consistent estimates of the solvent shifts are obtained as the difference between PCM-CAS(14,12)-srPBE values using the PCM geometry and the CAS(14,12)-srPBE values using the vacuum geometry.
However, previous theoretical studies of \ce{MnO4-} have occasionally neglected the indirect geometry effect of the solvent.\cite{Neugebauer_2005,Seth_2004}
The shift of $-0.14$ eV for the first intense transition reported in ref.~\citenum{Neugebauer_2005} is based on a vacuum excitation energy that was obtained from a structure optimized in vacuum (BPW91/TZ2P), and the solvated excitation energy was from the work by \citeauthor{Seth_2004}\cite{Seth_2004} where COSMO-SAOP/TZ2P was used on a structure that was also optimized in vacuum (BP86/TZ2P).
Meanwhile, the small shift (0.01 eV) reported by \citeauthor{Su_2013}\cite{Su_2013} was obtained with SAOP/TZ2P and COSMO.
If both structure and excitation energy are obtained consistently, then our estimate of the solvent shift is only $-0.01$ eV, which is in good agreement with the shift obtained by \citeauthor{Su_2013}\cite{Su_2013} but differs somewhat from the red-shift of $-0.14$ eV.
The discrepancy can, at least partly, be explained by the neglect of indirect solvent effects since we also obtain a small red-shift of $-0.05$ eV when using structures optimized in vacuum (see table~\ref{tbl:geomopt}).
All the same, even with a consistent procedure, implicit solvent models alone do not seem to be able to reproduce the experimental blue-shift which is around 0.1-0.2 eV.\cite{Houm_ller_2013}

\begin{table}
 \small
 \caption{Vertical excitation energies in eV and oscillator strengths of the twelve lowest intense transitions of the \ce{MnO4-} ion in vacuum or aqueous solution calculated as averages using structures from a molecular dynamics trajectory$^{a}$}
 \label{tbl:snapshots}
 \begin{tabular}{lccccccccc}
 \hline
  Geom. & & \multicolumn{2}{c}{QM MD$^{b}$} & \multicolumn{6}{c}{QM/MM MD$^{c}$} \\
  Environment & & \multicolumn{2}{c}{vacuum$^{d}$} & \multicolumn{2}{c}{vacuum$^{e}$} & \multicolumn{2}{c}{PCM$^{f}$} & \multicolumn{2}{c}{PE$^{g}$} \\
  \hline
  Multiplet&  State & $E_{\mathrm{vac}}$ & $f$ & $E_{\mathrm{aq}}$ & $f$ & $E_{\mathrm{aq}}$ & $f$ & $E_{\mathrm{aq}}$ & $f$ \\
  \hline
           &  1 & 1.93 & 0.013 & 1.97 & 0.008 & 1.98 & 0.011 & 1.91 & 0.010 \\
  1$^1$T$_1$   &  2 & 2.03 & 0.005 & 2.07 & 0.005 &  2.07 & 0.008 & 2.04 & 0.003 \\
           &  3 & 2.08 & 0.006 & 2.13 & 0.007 &  2.13 & 0.008 & 2.14 & 0.006 \\
  \hline
           &  4 & 2.19 & 0.014 & 2.23 & 0.017 &  2.22 & 0.015 & 2.30 & 0.016 \\
  1$^1$T$_2$   &  5 & 2.29 & 0.019 & 2.34 & 0.018 &  2.32 & 0.019 & 2.45 & 0.014 \\
           &  6 & 2.41 & 0.016 & 2.46 & 0.016 &  2.44 & 0.019 & 2.55 & 0.012 \\
  \hline    
           &  7 & 3.36 & 0.002 & - & - &  3.42 & 0.003 & 3.33 & 0.003 \\
           &  8 & 3.39 & 0.002 & - & - &  3.45 & 0.002 & 3.40 & 0.003 \\
2$^1$T$_1$/2$^1$T$_2$&  9 & 3.42 & 0.002 & - & - &  3.48 & 0.002 & 3.45 & 0.003 \\
           & 10 & 3.46 & 0.002 & - & - &  3.52 & 0.002 & 3.50 & 0.002 \\
           & 11 & 3.49 & 0.002 & - & - &  3.56 & 0.002 & 3.55 & 0.003 \\
           & 12 & 3.54 & 0.004 & - & - &  3.61 & 0.006 & 3.61 & 0.004 \\
  \hline
 \end{tabular} \\
 $^a$ \footnotesize{Calculated using a development version of Dalton\cite{daltonpaper,dalton_version} and PElib\cite{pelib} with (PCM-/PE-)CAS(14,12)-srPBE/cc-pVDZ with the srPBE(HSE;RI) exchange-correlation functional\cite{Heyd_2003} as defined in ref.~\citenum{Fromager_2007}.}	
 $^b$ \footnotesize{Averages based on structures taken from 1.0 ns BLYP/6-31G MD trajectory\cite{Olsen2015}.}
 $^c$ \footnotesize{Averages based on structures taken from ten 0.1 ns BLYP/6-31G//TIP3P MD trajectories\cite{Olsen2015}.}
 $^d$ \footnotesize{Averages based on 100 structures.}
 $^e$ \footnotesize{Averages based on 87 structures.}
 $^f$ \footnotesize{Cavity was built using 1.9 and 1.52 {\AA} for Mn and O atomic radii, respectively. Averages based on 87 structures.}
 $^g$ \footnotesize{Embedding potential based on B3LYP/cc-pVDZ calculations. Averages based on 99 structures.}
\end{table}

\begin{figure}[h]
 \includegraphics[width=0.5\textwidth]{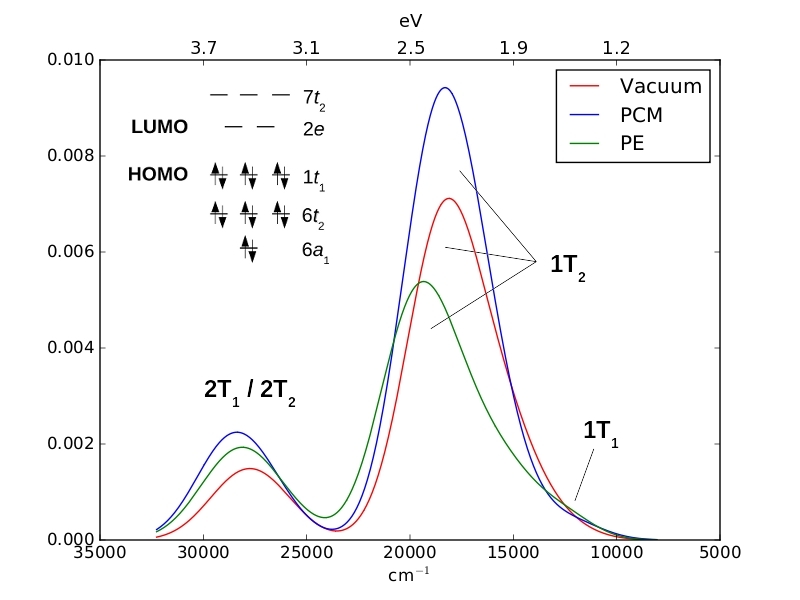}
 \caption{Simulated spectra of the \ce{MnO4-} ion in vacuum and aqueous solution modeled by the PCM or the PE model. The insert shows the generally accepted frontier molecular orbital diagram labeled according to the T$_d$ point group.}
 \label{fgr:spectra}
\end{figure}

We now consider \ce{MnO4-} in an explicit solvent environment.
We computed absorption spectra based on the twelve lowest states and in table~\ref{tbl:snapshots} we report $E_{\mathrm{vac}}$ and $E_{\mathrm{sol}}$ for these twelve states as averages based on structures taken from MD trajectories\cite{Olsen2015}.
Gaussian convolution of these states leads to the spectrum shown in figure~\ref{fgr:spectra}.
The two intense peaks in this figure correspond to the two lowest intense bands of $^1$T$_2$ parentage (corresponding to 1$^1$T$_2$ and 2$^1$T$_2$ for the symmetric structures in table~\ref{tbl:geomopt}).
The perturbation induced by the solvent molecules as well as the finite-temperature effects remove the degeneracy of the T$_1$ and T$_2$ states.
Thus, the T$_1$ and T$_2$ states are split into six states in total.
Further, the transition to the T$_1$ state is dipole-forbidden in T$_d$ symmetry, but gains intensity when the symmetry is removed by finite-temperature effects and solvent interactions.  
The peak-maxima in figure~\ref{fgr:spectra} are used to calculate $\Delta E_{\mathrm{shift}}$ which are given in table~\ref{tbl:shifts}.
We start by discussing these, and thereafter we will provide a more detailed analysis of the individual states in table~\ref{tbl:snapshots}.

From the spectrum we obtain an excitation energy of 2.25 eV for the 1$^1$T$_2$ state in vacuum.
This is 0.04 eV higher than the energy obtained using the geometry-optimized structure (see table~\ref{tbl:geomopt}) which can be attributed to finite-temperature effects in the ground-state.
The PCM predicts a very small blue-shift of 0.02 eV (see table~\ref{tbl:shifts}) and is thus similar to the shift obtained from the geometry-optimized structures.
The PE model predicts a more significant blue-shift of 0.15 eV (see table~\ref{tbl:shifts}) which is in excellent agreement with the experimental shift.

\begin{table}[h]
 \small
 \caption{Absorption maxima and solvent shifts in eV of the two lowest bands of \ce{MnO4-} ion obtained from the spectrum in figure~\ref{fgr:spectra}}
 \label{tbl:shifts}
 \begin{tabular}{cccccc}
  \hline
  Band & $E_{\mathrm{vac}}$ & $E^{\mathrm{PCM}}_{\mathrm{aq}}$ & $E^{\mathrm{PE}}_{\mathrm{aq}}$ & $E^{\mathrm{exp}}_{\mathrm{vac}}$\cite{Houm_ller_2013} & $E^{\mathrm{exp}}_{\mathrm{aq}}$\cite{Houm_ller_2013} \\
  \hline
  1st & 2.25 & 2.27 & 2.40 & 2.09/2.18 & 2.27/2.35 \\
  2nd & 3.44 & 3.52 & 3.49 & 3.55 & - \\
  \hline
   &  & $\Delta E^{\mathrm{PCM}}_{\mathrm{shift}}$ & $\Delta E^{\mathrm{PE}}_{\mathrm{shift}}$ & & $\Delta E^{\mathrm{exp}}_{\mathrm{shift}}$ \\
  \hline
  1st & - & 0.02 & 0.15 & - & 0.17/0.18 \\
  2nd & - & 0.08 & 0.05 & - & - \\
  \hline
 \end{tabular}
\end{table}

The spectrum in figure~\ref{fgr:spectra} can be interpreted on the basis of the individual states given in table~\ref{tbl:snapshots}.
We can compare these to the results obtained with a symmetric structure in table~\ref{tbl:geomopt}.
For the symmetric structure used in table~\ref{tbl:geomopt} we obtain an excitation energy of 2.13 eV for the 1$^1$T$_1$ multiplet (not shown in table~\ref{tbl:geomopt}), and with the excitation energy of 2.21 eV for the 1$^1$T$_2$ state (table~\ref{tbl:geomopt}), the splitting of the 1$^1$T$_1$ and 1$^1$T$_2$ states in vacuum is only 0.11 eV, similar  to results with RASPT2\cite{Su_2013}.
With such a small energy-gap, the two states are expected to mix when the tetrahedral symmetry is lifted. 
This is indeed the case, but among the lowest six states, the 1$^1$T$_1$ and 1$^1$T$_2$ multiplets can still be discerned (see~table~\ref{tbl:snapshots}).
The three states of 1$^1$T$_1$ parentage are still lower in energy than the intense 1$^1$T$_2$ band, and can also be discerned in the spectrum in figure~\ref{fgr:spectra}, as a weak transition at approximately 1.7-1.9 eV (13700-15300 cm$^{-1}$) prior to the large peak from 1$^1$T$_2$.
Low-intensity peaks have indeed previously 
been observed in \ce{MnO4-} spectra around 1.8 eV.\cite{Holt_1967,Houm_ller_2013}
Analysis have suggested that their origin were either in spin-forbidden $^3$T$_2$, orbital-forbidden $^1$T$_1$ states, or both.\cite{Ballhausen1963,Viste1964,Day1970}. 
From our calculations here, we cannot exclude the involvement of triplet states, but our results show that the $^1$T$_1$ states are likely to gain sufficient intensity to be responsible for the observed low-intensity band.
This assignment can naturally not be extracted from previous studies\cite{Seth_2004,Neugebauer_2005,Su_2013} with symmetric structures, 
where the intensity of the T$_1$ states are identically zero.

From inspection of table~\ref{tbl:snapshots} we also observe that the  direct electrostatic effect of the solvent is predicted different by the PE model compared to the PCM. 
In both cases, the indirect solvent effect, coming from the change in geometry of the solvated \ce{MnO4-} ion, results in a blue-shift of the excitation energies.
This can be seen in table~\ref{tbl:snapshots} by comparing the excitation energies based on QM MD geometries to QM/MM MD geometries both with vacuum environment.
For the PCM the direct solvent effects (obtained from column 5 and 7 in table~\ref{tbl:snapshots}) is a (small) red-shift of the excitation energies.
Adding this to the indirect solvent shift thus gives the very small blue-shift as discussed above.
Meanwhile, the direct solvent effects modeled by the PE model (see table~\ref{tbl:snapshots}) result in a blue-shift of the excitation energy, leading to a more pronounced total blue-shift.

We can also briefly comment on the higher-lying states in table~\ref{tbl:snapshots} and figure~\ref{fgr:spectra}. 
Our CAS-srPBE results from the symmetric structure in table~\ref{tbl:geomopt} showed that the dipole-allowed 2$^1$T$_2$ multiplet is just above the dipole-forbidden 2$^1$T$_1$ multiplet: 
The excitation energy of the former is 3.61 eV (see table~\ref{tbl:geomopt}) whereas the 2$^1$T$_1$ multiplet has an excitation energy of 3.48 eV (not shown in table~\ref{tbl:geomopt}).
The RAS(24,17)PT2 results from ref.~\citenum{Su_2013} predict a similar situation where the excitation energies are 3.53 eV (2$^1$T$_1$) and 3.39 eV (2$^1$T$_1$), respectively.
These two multiplets are mixed when including finite-temperature and solvent effects and can no longer be discerned (see table~\ref{tbl:snapshots}). 
From the spectra in figure~\ref{fgr:spectra} we obtain an excitation energy of 3.44 eV in vacuum, and blue-shifts of 0.08 and 0.05 for the PCM and the PE model, respectively 
(see table~\ref{tbl:shifts}). There is no experimental value for this transition in aqueous solution but a value of 0.09 eV exists for a 0-0 transition in crystalline phase.\cite{Houm_ller_2013,Holt_1967} Using this value as reference, both solvent models yield a qualitatively reasonable shift (see table~\ref{tbl:shifts}), in particular in the light that the shift of excitation energies due to an environment seem to be larger in the crystalline phase than in water.\cite{Houm_ller_2013} Our calculated shifts for water are indeed slightly below the value for the crystalline phase.

It should finally be noted that we do not reproduce the vibrational structure of the spectrum, since our underlying structures only include vibrations from the electronic ground state, while the significant vibrational structure in the \ce{MnO4-} spectrum is a result of vibrational coupling in the excited states.
The correct description of this vibronic coupling was a notable achievement of ref.~\citenum{Neugebauer_2005}. Although the authors note that the absolute excitation energies are significantly overestimated by the TD-DFT, they also find that the method convincingly reproduces the vibrational structure through an effective Hamiltonian approach.
Employing a similar model for CAS-srDFT would be an  interesting extension of our current approach but is left for further studies.
In any case, the good agreement with experiment obtained here is encouraging for further investigating 
the PE-CAS-srDFT method for other transition metals, both in solution and in protein systems.

\section{Conclusion}

In this paper, we present a study of \ce{MnO4-} in vacuum and in aqueous solution, employing the multiconfigurational CAS-srDFT method combined with implicit and explicit solvation models. In vacuum calculations, we show that CAS-srDFT is at least as accurate as RASPT2, although we can employ significantly smaller active spaces with CAS-srDFT.
To address the solvent shift, we employed a combination of CAS-srDFT and the PE model (explicit solvation), as well as the PCM (implicit solvation).
Finite-temperature effects are included by using structures obtained from QM and QM/MM MD trajectories.
An experimental study has predicted that the lowest intense transition is blue-shifted by about 0.1-0.2 eV compared to vacuum.
However, previous theoretical studies that employed continuum solvation models predicted that the solvent shift is essentially zero.
We find that the reason lies in an unsatisfactory description of the direct electrostatic interactions between solute and solvent.
The solvent-induced perturbation of the solute geometry shifts the excitation energy about 0.05 eV.
Adding the direct electrostatic interactions from the solvent via PCM red-shifts the energy by about 0.03 eV, leading to a total blue-shift of only 0.02 eV.
Using the PE model instead adds an additional blue-shift of about 0.1 eV, so that the total shift is 0.15 eV.
Thus the combined use of the CAS-srDFT and the PE model yield both a solvent shift and absolute excitation energies that are in excellent agreement with experiment.

\begin{acknowledgement}

J.~M.~H.~O.\ acknowledges financial support from the Danish Council for Independent Research (DFF) through the Sapere Aude research career program (grant id.~DFF -- 1325-00091 and DFF -- 1323-00744) and the Carlsberg Foundation (grant id.~CF15-0823).
E.~D.~H.\ thanks the Carlsberg Foundation (grant id.~CF16-0482) and the Villum Foundation for funding.

\end{acknowledgement}

\bibliography{manuscript}

\providecommand{\latin}[1]{#1}
\providecommand*\mcitethebibliography{\thebibliography}
\csname @ifundefined\endcsname{endmcitethebibliography}
  {\let\endmcitethebibliography\endthebibliography}{}
\begin{mcitethebibliography}{53}
\providecommand*\natexlab[1]{#1}
\providecommand*\mciteSetBstSublistMode[1]{}
\providecommand*\mciteSetBstMaxWidthForm[2]{}
\providecommand*\mciteBstWouldAddEndPuncttrue
  {\def\EndOfBibitem{\unskip.}}
\providecommand*\mciteBstWouldAddEndPunctfalse
  {\let\EndOfBibitem\relax}
\providecommand*\mciteSetBstMidEndSepPunct[3]{}
\providecommand*\mciteSetBstSublistLabelBeginEnd[3]{}
\providecommand*\EndOfBibitem{}
\mciteSetBstSublistMode{f}
\mciteSetBstMaxWidthForm{subitem}{(\alph{mcitesubitemcount})}
\mciteSetBstSublistLabelBeginEnd
  {\mcitemaxwidthsubitemform\space}
  {\relax}
  {\relax}

\bibitem[Andersson \latin{et~al.}(1992)Andersson, Malmqvist, and
  Roos]{Andersson1992}
Andersson,~K.; Malmqvist,~P.; Roos,~B.~O. \emph{J. Chem. Phys.} \textbf{1992},
  \emph{96}, 1218--1226\relax
\mciteBstWouldAddEndPuncttrue
\mciteSetBstMidEndSepPunct{\mcitedefaultmidpunct}
{\mcitedefaultendpunct}{\mcitedefaultseppunct}\relax
\EndOfBibitem
\bibitem[Wolfsberg and Helmholz(1952)Wolfsberg, and Helmholz]{wolfsberg1952}
Wolfsberg,~M.; Helmholz,~L. \emph{{J. Chem. Phys.}} \textbf{1952}, \emph{20},
  837--843\relax
\mciteBstWouldAddEndPuncttrue
\mciteSetBstMidEndSepPunct{\mcitedefaultmidpunct}
{\mcitedefaultendpunct}{\mcitedefaultseppunct}\relax
\EndOfBibitem
\bibitem[Johnson and Smith(1971)Johnson, and Smith]{Johnson_1971}
Johnson,~K.; Smith,~F.~J. \emph{{Chem. Phys. Lett.}} \textbf{1971}, \emph{10},
  219--223\relax
\mciteBstWouldAddEndPuncttrue
\mciteSetBstMidEndSepPunct{\mcitedefaultmidpunct}
{\mcitedefaultendpunct}{\mcitedefaultseppunct}\relax
\EndOfBibitem
\bibitem[Johansen and Rettrup(1983)Johansen, and Rettrup]{Johansen_1983}
Johansen,~H.; Rettrup,~S. \emph{{Chem. Phys.}} \textbf{1983}, \emph{74},
  77--81\relax
\mciteBstWouldAddEndPuncttrue
\mciteSetBstMidEndSepPunct{\mcitedefaultmidpunct}
{\mcitedefaultendpunct}{\mcitedefaultseppunct}\relax
\EndOfBibitem
\bibitem[Nakai \latin{et~al.}(1991)Nakai, Ohmori, and Nakatsuji]{Nakai_1991}
Nakai,~H.; Ohmori,~Y.; Nakatsuji,~H. \emph{J. Chem. Phys.} \textbf{1991},
  \emph{95}, 8287--8291\relax
\mciteBstWouldAddEndPuncttrue
\mciteSetBstMidEndSepPunct{\mcitedefaultmidpunct}
{\mcitedefaultendpunct}{\mcitedefaultseppunct}\relax
\EndOfBibitem
\bibitem[Nooijen(1999)]{Nooijen_1999}
Nooijen,~M. \emph{J. Chem. Phys.} \textbf{1999}, \emph{111}, 10815--10826\relax
\mciteBstWouldAddEndPuncttrue
\mciteSetBstMidEndSepPunct{\mcitedefaultmidpunct}
{\mcitedefaultendpunct}{\mcitedefaultseppunct}\relax
\EndOfBibitem
\bibitem[van Gisbergen \latin{et~al.}(1999)van Gisbergen, Groeneveld, Rosa,
  Snijders, and Baerends]{Gisbergen_1999}
van Gisbergen,~S. J.~A.; Groeneveld,~J.~A.; Rosa,~A.; Snijders,~J.~G.;
  Baerends,~E.~J. \emph{{J. Phys. Chem. A}} \textbf{1999}, \emph{103},
  6835--6844\relax
\mciteBstWouldAddEndPuncttrue
\mciteSetBstMidEndSepPunct{\mcitedefaultmidpunct}
{\mcitedefaultendpunct}{\mcitedefaultseppunct}\relax
\EndOfBibitem
\bibitem[Nooijen and Lotrich(2000)Nooijen, and Lotrich]{Nooijen_2000}
Nooijen,~M.; Lotrich,~V. \emph{{J. Chem. Phys.}} \textbf{2000}, \emph{113},
  494--507\relax
\mciteBstWouldAddEndPuncttrue
\mciteSetBstMidEndSepPunct{\mcitedefaultmidpunct}
{\mcitedefaultendpunct}{\mcitedefaultseppunct}\relax
\EndOfBibitem
\bibitem[Boulet \latin{et~al.}(2001)Boulet, Chermette, Daul, Gilardoni,
  Rogemond, Weber, and Zuber]{Boulet_2001}
Boulet,~P.; Chermette,~H.; Daul,~C.; Gilardoni,~F.; Rogemond,~F.; Weber,~J.;
  Zuber,~G. \emph{{J. Phys. Chem. A}} \textbf{2001}, \emph{105}, 885--894\relax
\mciteBstWouldAddEndPuncttrue
\mciteSetBstMidEndSepPunct{\mcitedefaultmidpunct}
{\mcitedefaultendpunct}{\mcitedefaultseppunct}\relax
\EndOfBibitem
\bibitem[Neugebauer \latin{et~al.}(2005)Neugebauer, Baerends, and
  Nooijen]{Neugebauer_2005}
Neugebauer,~J.; Baerends,~E.~J.; Nooijen,~M. \emph{{J. Phys. Chem. A}}
  \textbf{2005}, \emph{109}, 1168--1179\relax
\mciteBstWouldAddEndPuncttrue
\mciteSetBstMidEndSepPunct{\mcitedefaultmidpunct}
{\mcitedefaultendpunct}{\mcitedefaultseppunct}\relax
\EndOfBibitem
\bibitem[Jose \latin{et~al.}(2012)Jose, Seth, and Ziegler]{Linta_2012}
Jose,~L.; Seth,~M.; Ziegler,~T. \emph{J. Phys. Chem. A} \textbf{2012},
  \emph{116}, 1864--1876\relax
\mciteBstWouldAddEndPuncttrue
\mciteSetBstMidEndSepPunct{\mcitedefaultmidpunct}
{\mcitedefaultendpunct}{\mcitedefaultseppunct}\relax
\EndOfBibitem
\bibitem[Ziegler(2012)]{Ziegler_2012}
Ziegler,~T. \emph{{Struct. Bond}} \textbf{2012}, \emph{143}, 1--38\relax
\mciteBstWouldAddEndPuncttrue
\mciteSetBstMidEndSepPunct{\mcitedefaultmidpunct}
{\mcitedefaultendpunct}{\mcitedefaultseppunct}\relax
\EndOfBibitem
\bibitem[Almeida \latin{et~al.}(2015)Almeida, McKinlay, and
  Paterson]{Almeida_2015}
Almeida,~N.~M.; McKinlay,~R.~G.; Paterson,~M.~J. \emph{Chem. Phys.}
  \textbf{2015}, \emph{446}, 86--91\relax
\mciteBstWouldAddEndPuncttrue
\mciteSetBstMidEndSepPunct{\mcitedefaultmidpunct}
{\mcitedefaultendpunct}{\mcitedefaultseppunct}\relax
\EndOfBibitem
\bibitem[Seidu \latin{et~al.}(2015)Seidu, Krykunov, and Ziegler]{Seidu_2015}
Seidu,~I.; Krykunov,~M.; Ziegler,~T. \emph{J. Chem. Theory Comput.}
  \textbf{2015}, \emph{11}, 4041--4053, PMID: 26575900\relax
\mciteBstWouldAddEndPuncttrue
\mciteSetBstMidEndSepPunct{\mcitedefaultmidpunct}
{\mcitedefaultendpunct}{\mcitedefaultseppunct}\relax
\EndOfBibitem
\bibitem[Buijse and Baerends(1990)Buijse, and Baerends]{Buijse_1990}
Buijse,~M.~A.; Baerends,~E.~J. \emph{{J. Chem. Phys.}} \textbf{1990},
  \emph{93}, 4129--4141\relax
\mciteBstWouldAddEndPuncttrue
\mciteSetBstMidEndSepPunct{\mcitedefaultmidpunct}
{\mcitedefaultendpunct}{\mcitedefaultseppunct}\relax
\EndOfBibitem
\bibitem[Holt and Ballhausen(1967)Holt, and Ballhausen]{Holt_1967}
Holt,~S.~L.; Ballhausen,~C. \emph{{Theor. Chim. Acta}} \textbf{1967}, \emph{7},
  313--320\relax
\mciteBstWouldAddEndPuncttrue
\mciteSetBstMidEndSepPunct{\mcitedefaultmidpunct}
{\mcitedefaultendpunct}{\mcitedefaultseppunct}\relax
\EndOfBibitem
\bibitem[Veryazov \latin{et~al.}(2011)Veryazov, Malmqvist, and
  Roos]{Veryazov_2011}
Veryazov,~V.; Malmqvist,~P.~Ã.; Roos,~B.~O. \emph{{Int. J. Quantum Chem.}}
  \textbf{2011}, \emph{111}, 3329--3338\relax
\mciteBstWouldAddEndPuncttrue
\mciteSetBstMidEndSepPunct{\mcitedefaultmidpunct}
{\mcitedefaultendpunct}{\mcitedefaultseppunct}\relax
\EndOfBibitem
\bibitem[Hedeg{\aa}rd \latin{et~al.}(2014)Hedeg{\aa}rd, Jensen, and
  Kongsted]{Hedegaard_2014}
Hedeg{\aa}rd,~E.~D.; Jensen,~H. J.~A.; Kongsted,~J. \emph{{Int. J. Quantum
  Chem.}} \textbf{2014}, \emph{114}, 1102--1107\relax
\mciteBstWouldAddEndPuncttrue
\mciteSetBstMidEndSepPunct{\mcitedefaultmidpunct}
{\mcitedefaultendpunct}{\mcitedefaultseppunct}\relax
\EndOfBibitem
\bibitem[Stein and Reiher(2016)Stein, and Reiher]{Stein_2016}
Stein,~C.~J.; Reiher,~M. \emph{J. Chem. Theory Comput.} \textbf{2016},
  \emph{12}, 1760--1771\relax
\mciteBstWouldAddEndPuncttrue
\mciteSetBstMidEndSepPunct{\mcitedefaultmidpunct}
{\mcitedefaultendpunct}{\mcitedefaultseppunct}\relax
\EndOfBibitem
\bibitem[Su \latin{et~al.}(2013)Su, Xu, Xu, Schwarz, and Li]{Su_2013}
Su,~J.; Xu,~W.-H.; Xu,~C.-F.; Schwarz,~W. H.~E.; Li,~J. \emph{Inorg. Chem.}
  \textbf{2013}, \emph{52}, 9867--9874\relax
\mciteBstWouldAddEndPuncttrue
\mciteSetBstMidEndSepPunct{\mcitedefaultmidpunct}
{\mcitedefaultendpunct}{\mcitedefaultseppunct}\relax
\EndOfBibitem
\bibitem[Houm{\o}ller \latin{et~al.}(2013)Houm{\o}ller, Kaufman, St{\o}chkel,
  Tribedi, Nielsen, and Weber]{Houm_ller_2013}
Houm{\o}ller,~J.; Kaufman,~S.~H.; St{\o}chkel,~K.; Tribedi,~L.~C.;
  Nielsen,~S.~B.; Weber,~J.~M. \emph{{ChemPhysChem}} \textbf{2013}, \emph{14},
  1133--1137\relax
\mciteBstWouldAddEndPuncttrue
\mciteSetBstMidEndSepPunct{\mcitedefaultmidpunct}
{\mcitedefaultendpunct}{\mcitedefaultseppunct}\relax
\EndOfBibitem
\bibitem[Seth \latin{et~al.}(2004)Seth, Ziegler, Banerjee, Autschbach, van
  Gisbergen, and Baerends]{Seth_2004}
Seth,~M.; Ziegler,~T.; Banerjee,~A.; Autschbach,~J.; van Gisbergen,~S. J.~A.;
  Baerends,~E.~J. \emph{J. Chem. Phys.} \textbf{2004}, \emph{120},
  10942--10954\relax
\mciteBstWouldAddEndPuncttrue
\mciteSetBstMidEndSepPunct{\mcitedefaultmidpunct}
{\mcitedefaultendpunct}{\mcitedefaultseppunct}\relax
\EndOfBibitem
\bibitem[Savin(1996)]{savinbook}
Savin,~A. \emph{{Recent Developments and Applications of Modern Density
  Functional Theory}}; Elsevier: Amsterdam, 1996; p 327\relax
\mciteBstWouldAddEndPuncttrue
\mciteSetBstMidEndSepPunct{\mcitedefaultmidpunct}
{\mcitedefaultendpunct}{\mcitedefaultseppunct}\relax
\EndOfBibitem
\bibitem[Savin and Flad(1995)Savin, and Flad]{Savin_1995}
Savin,~A.; Flad,~H.-J. \emph{Int. J. Quantum Chem.} \textbf{1995}, \emph{56},
  327--332\relax
\mciteBstWouldAddEndPuncttrue
\mciteSetBstMidEndSepPunct{\mcitedefaultmidpunct}
{\mcitedefaultendpunct}{\mcitedefaultseppunct}\relax
\EndOfBibitem
\bibitem[Leininger \latin{et~al.}(1997)Leininger, Stoll, Werner, and
  Savin]{leininger1997}
Leininger,~T.; Stoll,~H.; Werner,~H.-J.; Savin,~A. \emph{Chem. Phys. Lett.}
  \textbf{1997}, \emph{275}, 151\relax
\mciteBstWouldAddEndPuncttrue
\mciteSetBstMidEndSepPunct{\mcitedefaultmidpunct}
{\mcitedefaultendpunct}{\mcitedefaultseppunct}\relax
\EndOfBibitem
\bibitem[\'Angy\'an \latin{et~al.}(2005)\'Angy\'an, Gerber, Savin, and
  Toulouse]{Angyan_2005}
\'Angy\'an,~J.~G.; Gerber,~I.~C.; Savin,~A.; Toulouse,~J. \emph{Phys. Rev. A}
  \textbf{2005}, \emph{72}, 012510\relax
\mciteBstWouldAddEndPuncttrue
\mciteSetBstMidEndSepPunct{\mcitedefaultmidpunct}
{\mcitedefaultendpunct}{\mcitedefaultseppunct}\relax
\EndOfBibitem
\bibitem[Fromager \latin{et~al.}(2007)Fromager, Toulouse, and
  Jensen]{Fromager_2007}
Fromager,~E.; Toulouse,~J.; Jensen,~H. J.~A. \emph{{J. Chem. Phys.}}
  \textbf{2007}, \emph{126}, 074111\relax
\mciteBstWouldAddEndPuncttrue
\mciteSetBstMidEndSepPunct{\mcitedefaultmidpunct}
{\mcitedefaultendpunct}{\mcitedefaultseppunct}\relax
\EndOfBibitem
\bibitem[Fromager and Jensen(2008)Fromager, and Jensen]{Fromager_2008}
Fromager,~E.; Jensen,~H. J.~A. \emph{Phys. Rev. A} \textbf{2008}, \emph{78},
  022504\relax
\mciteBstWouldAddEndPuncttrue
\mciteSetBstMidEndSepPunct{\mcitedefaultmidpunct}
{\mcitedefaultendpunct}{\mcitedefaultseppunct}\relax
\EndOfBibitem
\bibitem[Fromager \latin{et~al.}(2010)Fromager, Cimiraglia, and
  Jensen]{Fromager_2010}
Fromager,~E.; Cimiraglia,~R.; Jensen,~H. J.~A. \emph{Phys. Rev. A}
  \textbf{2010}, \emph{81}, 024502\relax
\mciteBstWouldAddEndPuncttrue
\mciteSetBstMidEndSepPunct{\mcitedefaultmidpunct}
{\mcitedefaultendpunct}{\mcitedefaultseppunct}\relax
\EndOfBibitem
\bibitem[Toulouse \latin{et~al.}(2009)Toulouse, Gerber, Jansen, Savin, and
  \'Angy\'an]{Toulouse_2009}
Toulouse,~J.; Gerber,~I.~C.; Jansen,~G.; Savin,~A.; \'Angy\'an,~J.~G.
  \emph{Phys. Rev. Lett.} \textbf{2009}, \emph{102}, 096404\relax
\mciteBstWouldAddEndPuncttrue
\mciteSetBstMidEndSepPunct{\mcitedefaultmidpunct}
{\mcitedefaultendpunct}{\mcitedefaultseppunct}\relax
\EndOfBibitem
\bibitem[Fromager \latin{et~al.}(2013)Fromager, Knecht, and
  Jensen]{Fromager_2013}
Fromager,~E.; Knecht,~S.; Jensen,~H. J.~A. \emph{{J. Chem. Phys.}}
  \textbf{2013}, \emph{138}, 084101\relax
\mciteBstWouldAddEndPuncttrue
\mciteSetBstMidEndSepPunct{\mcitedefaultmidpunct}
{\mcitedefaultendpunct}{\mcitedefaultseppunct}\relax
\EndOfBibitem
\bibitem[Hedeg{\aa}rd \latin{et~al.}(2013)Hedeg{\aa}rd, Heiden, Knecht,
  Fromager, and Jensen]{Hedegaard_2013b}
Hedeg{\aa}rd,~E.~D.; Heiden,~F.; Knecht,~S.; Fromager,~E.; Jensen,~H. J.~A.
  \emph{{J. Chem. Phys.}} \textbf{2013}, \emph{139}, 184308\relax
\mciteBstWouldAddEndPuncttrue
\mciteSetBstMidEndSepPunct{\mcitedefaultmidpunct}
{\mcitedefaultendpunct}{\mcitedefaultseppunct}\relax
\EndOfBibitem
\bibitem[Hedeg{\aa}rd \latin{et~al.}(2015)Hedeg{\aa}rd, Knecht, Kielberg,
  Jensen, and Reiher]{Hedegaard_2015b}
Hedeg{\aa}rd,~E.~D.; Knecht,~S.; Kielberg,~J.~S.; Jensen,~H. J.~A.; Reiher,~M.
  \emph{J. Chem. Phys.} \textbf{2015}, \emph{142}, 224108\relax
\mciteBstWouldAddEndPuncttrue
\mciteSetBstMidEndSepPunct{\mcitedefaultmidpunct}
{\mcitedefaultendpunct}{\mcitedefaultseppunct}\relax
\EndOfBibitem
\bibitem[Hedeg{\aa}rd()]{Hedegaard_2016a}
Hedeg{\aa}rd,~E.~D. \emph{Mol. Phys.} \relax
\mciteBstWouldAddEndPunctfalse
\mciteSetBstMidEndSepPunct{\mcitedefaultmidpunct}
{}{\mcitedefaultseppunct}\relax
\EndOfBibitem
\bibitem[Hubert \latin{et~al.}(2016)Hubert, Hedeg{\aa}rd, and
  Jensen]{Hubert_2016a}
Hubert,~M.; Hedeg{\aa}rd,~E.~D.; Jensen,~H. J.~A. \emph{J. Chem. Theory
  Comput.} \textbf{2016}, \emph{12}, 2203--2213, PMID: 27058733\relax
\mciteBstWouldAddEndPuncttrue
\mciteSetBstMidEndSepPunct{\mcitedefaultmidpunct}
{\mcitedefaultendpunct}{\mcitedefaultseppunct}\relax
\EndOfBibitem
\bibitem[Hubert \latin{et~al.}(2016)Hubert, Jensen, and
  Hedeg{\aa}rd]{Hubert_2016b}
Hubert,~M.; Jensen,~H. J.~A.; Hedeg{\aa}rd,~E.~D. \emph{J. Phys. Chem. A}
  \textbf{2016}, \emph{120}, 36--43, PMID: 26669578\relax
\mciteBstWouldAddEndPuncttrue
\mciteSetBstMidEndSepPunct{\mcitedefaultmidpunct}
{\mcitedefaultendpunct}{\mcitedefaultseppunct}\relax
\EndOfBibitem
\bibitem[Senjean \latin{et~al.}(2016)Senjean, Hedeg{\aa}rd, Alam, Knecht, and
  Fromager]{Senjean_2016}
Senjean,~B.; Hedeg{\aa}rd,~E.~D.; Alam,~M.~M.; Knecht,~S.; Fromager,~E.
  \emph{Mol. Phys.} \textbf{2016}, \emph{114}, 968--981\relax
\mciteBstWouldAddEndPuncttrue
\mciteSetBstMidEndSepPunct{\mcitedefaultmidpunct}
{\mcitedefaultendpunct}{\mcitedefaultseppunct}\relax
\EndOfBibitem
\bibitem[Grimme and Waletzke(1999)Grimme, and Waletzke]{Grimme_1999}
Grimme,~S.; Waletzke,~M. \emph{J. Chem. Phys.} \textbf{1999}, \emph{111},
  5645--5655\relax
\mciteBstWouldAddEndPuncttrue
\mciteSetBstMidEndSepPunct{\mcitedefaultmidpunct}
{\mcitedefaultendpunct}{\mcitedefaultseppunct}\relax
\EndOfBibitem
\bibitem[Marian and Gilka(2008)Marian, and Gilka]{Marian_2008}
Marian,~C.~M.; Gilka,~N. \emph{J. Chem. Theory Comput.} \textbf{2008},
  \emph{4}, 1501--1515\relax
\mciteBstWouldAddEndPuncttrue
\mciteSetBstMidEndSepPunct{\mcitedefaultmidpunct}
{\mcitedefaultendpunct}{\mcitedefaultseppunct}\relax
\EndOfBibitem
\bibitem[Manni \latin{et~al.}(2014)Manni, Carlson, Luo, Ma, Olsen, Truhlar, and
  Gagliardi]{Manni_2014}
Manni,~G.~L.; Carlson,~R.~K.; Luo,~S.; Ma,~D.; Olsen,~J.; Truhlar,~D.~G.;
  Gagliardi,~L. \emph{J. Chem. Theory Comput.} \textbf{2014}, \emph{10},
  3669--3680\relax
\mciteBstWouldAddEndPuncttrue
\mciteSetBstMidEndSepPunct{\mcitedefaultmidpunct}
{\mcitedefaultendpunct}{\mcitedefaultseppunct}\relax
\EndOfBibitem
\bibitem[Garza \latin{et~al.}(2015)Garza, Bulik, Henderson, and
  Scuseria]{Garza_2015}
Garza,~A.~J.; Bulik,~I.~W.; Henderson,~T.~M.; Scuseria,~G.~E. \emph{Phys. Chem.
  Chem. Phys.} \textbf{2015}, \emph{17}, 22412--22422\relax
\mciteBstWouldAddEndPuncttrue
\mciteSetBstMidEndSepPunct{\mcitedefaultmidpunct}
{\mcitedefaultendpunct}{\mcitedefaultseppunct}\relax
\EndOfBibitem
\bibitem[Olsen \latin{et~al.}(2010)Olsen, Aidas, and Kongsted]{Olsen_2010}
Olsen,~J.~M.; Aidas,~K.; Kongsted,~J. \emph{J. Chem. Theory Comput.}
  \textbf{2010}, \emph{6}, 3721--3734\relax
\mciteBstWouldAddEndPuncttrue
\mciteSetBstMidEndSepPunct{\mcitedefaultmidpunct}
{\mcitedefaultendpunct}{\mcitedefaultseppunct}\relax
\EndOfBibitem
\bibitem[Olsen and Kongsted(2011)Olsen, and Kongsted]{Olsen_2011}
Olsen,~J.~M.; Kongsted,~J. \emph{Adv. Quantum Chem.} \textbf{2011}, \emph{61},
  107--143\relax
\mciteBstWouldAddEndPuncttrue
\mciteSetBstMidEndSepPunct{\mcitedefaultmidpunct}
{\mcitedefaultendpunct}{\mcitedefaultseppunct}\relax
\EndOfBibitem
\bibitem[Hedeg{\aa}rd \latin{et~al.}(2015)Hedeg{\aa}rd, Olsen, Knecht,
  Kongsted, and Jensen]{Hedegaard_2015a}
Hedeg{\aa}rd,~E.~D.; Olsen,~J. M.~H.; Knecht,~S.; Kongsted,~J.; Jensen,~H.
  J.~A. \emph{{J. Chem. Phys.}} \textbf{2015}, \emph{142}, 114113\relax
\mciteBstWouldAddEndPuncttrue
\mciteSetBstMidEndSepPunct{\mcitedefaultmidpunct}
{\mcitedefaultendpunct}{\mcitedefaultseppunct}\relax
\EndOfBibitem
\bibitem[Aidas \latin{et~al.}(2014)Aidas, Angeli, Bak, Bakken, Bast, Boman,
  Christiansen, Cimiraglia, Coriani, Dahle, Dalskov, Ekstr\"{o}m, Enevoldsen,
  Eriksen, Ettenhuber, Fern\'{a}ndez, Ferrighi, Fliegl, Frediani, Hald,
  Halkier, H\"{a}ttig, Heiberg, Helgaker, Hennum, Hettema, Hjerten\ae{}s,
  H\o{}st, H\o{}yvik, Iozzi, Jans\'{i}k, Jensen, Jonsson, J\o{}rgensen,
  Kauczor, Kirpekar, Kj\ae{}rgaard, Klopper, Knecht, Kobayashi, Koch, Kongsted,
  Krapp, Kristensen, Ligabue, Lutn\ae{}s, Melo, Mikkelsen, Myhre, Neiss,
  Nielsen, Norman, Olsen, Olsen, Osted, Packer, Pawlowski, Pedersen, Provasi,
  Reine, Rinkevicius, Ruden, Ruud, Rybkin, Sa\l{}ek, Samson, de~Mer\'{a}s,
  Saue, Sauer, Schimmelpfennig, Sneskov, Steindal, Sylvester-Hvid, Taylor,
  Teale, Tellgren, Tew, Thorvaldsen, Th\o{}gersen, Vahtras, Watson, Wilson,
  Ziolkowski, and \AA{}gren]{daltonpaper}
Aidas,~K. \latin{et~al.}  \emph{WIREs Comput.~Mol.~Sci.} \textbf{2014},
  \emph{4}, 269--284\relax
\mciteBstWouldAddEndPuncttrue
\mciteSetBstMidEndSepPunct{\mcitedefaultmidpunct}
{\mcitedefaultendpunct}{\mcitedefaultseppunct}\relax
\EndOfBibitem
\bibitem[dal()]{dalton_version}
{Dalton}, {a Molecular Electronic Structure Program}, development version, see
  \texttt{http://daltonprogram.org/}\relax
\mciteBstWouldAddEndPuncttrue
\mciteSetBstMidEndSepPunct{\mcitedefaultmidpunct}
{\mcitedefaultendpunct}{\mcitedefaultseppunct}\relax
\EndOfBibitem
\bibitem[Heyd \latin{et~al.}(2003)Heyd, Scuseria, and Ernzerhof]{Heyd_2003}
Heyd,~J.; Scuseria,~G.~E.; Ernzerhof,~M. \emph{J. Chem. Phys.} \textbf{2003},
  \emph{118}, 8207--8215\relax
\mciteBstWouldAddEndPuncttrue
\mciteSetBstMidEndSepPunct{\mcitedefaultmidpunct}
{\mcitedefaultendpunct}{\mcitedefaultseppunct}\relax
\EndOfBibitem
\bibitem[{The PE library developers}(2016)]{pelib}
{The PE library developers}, {PElib: The Polarizable Embedding library (version
  1.2.4)}. 2016; https://gitlab.com/pe-software/pelib-public\relax
\mciteBstWouldAddEndPuncttrue
\mciteSetBstMidEndSepPunct{\mcitedefaultmidpunct}
{\mcitedefaultendpunct}{\mcitedefaultseppunct}\relax
\EndOfBibitem
\bibitem[Olsen \latin{et~al.}(2015)Olsen, Steinmann, Ruud, and
  Kongsted]{Olsen2015}
Olsen,~J. M.~H.; Steinmann,~C.; Ruud,~K.; Kongsted,~J. \emph{J. Phys. Chem. A}
  \textbf{2015}, \emph{119}, 5344--5355\relax
\mciteBstWouldAddEndPuncttrue
\mciteSetBstMidEndSepPunct{\mcitedefaultmidpunct}
{\mcitedefaultendpunct}{\mcitedefaultseppunct}\relax
\EndOfBibitem
\bibitem[Ballhausen(1963)]{Ballhausen1963}
Ballhausen,~C.~J. \emph{Theor. Chim. Acta} \textbf{1963}, \emph{1},
  285--293\relax
\mciteBstWouldAddEndPuncttrue
\mciteSetBstMidEndSepPunct{\mcitedefaultmidpunct}
{\mcitedefaultendpunct}{\mcitedefaultseppunct}\relax
\EndOfBibitem
\bibitem[Viste and Gray(1964)Viste, and Gray]{Viste1964}
Viste,~A.; Gray,~H.~B. \emph{Inorg. Chem.} \textbf{1964}, \emph{3},
  1113--1123\relax
\mciteBstWouldAddEndPuncttrue
\mciteSetBstMidEndSepPunct{\mcitedefaultmidpunct}
{\mcitedefaultendpunct}{\mcitedefaultseppunct}\relax
\EndOfBibitem
\bibitem[Day \latin{et~al.}(1970)Day, Disipio, and Oleari]{Day1970}
Day,~P.; Disipio,~L.; Oleari,~L. \emph{Chem. Phys. Lett.} \textbf{1970},
  \emph{5}, 533 -- 536\relax
\mciteBstWouldAddEndPuncttrue
\mciteSetBstMidEndSepPunct{\mcitedefaultmidpunct}
{\mcitedefaultendpunct}{\mcitedefaultseppunct}\relax
\EndOfBibitem
\end{mcitethebibliography}

\end{document}